# Evolution of diverse (and advanced) cognitive abilities through adaptive fine-tuning of learning and chunking mechanisms

Arnon Lotem[1*] and Joseph Y. Halpern[2]

1. School of Zoology, Faculty of Life Sciences and Sagol School of Neurosciences, Tel-Aviv University, Tel-Aviv, Israel.  E-mail: lotem@tauex.tau.ac.il,  ORCID ID: https://orcid.org/0000-0001-5428-2005

2. Dept. of Computer Science, Cornell University, Ithaca NY 14850, USA. E-mail: halpern@cs.cornell.edu

* Corresponding author

Abstract

The evolution of cognition is frequently discussed as the evolution of cognitive abilities or the evolution of some neuronal structures in the brain. However, since such traits or abilities are often highly complex, understanding their evolution requires explaining how they could have gradually evolved through selection acting on heritable variations in simpler cognitive mechanisms. With this in mind, making use of a previously proposed theory, here we show how the evolution of cognitive abilities can be captured by the fine-tuning of basic learning mechanisms and, in particular, chunking mechanisms. We use the term *chunking* broadly for all types of non-elemental learning, claiming that the process by which elements are combined into chunks and associated with other chunks, or elements, is critical for what the brain can do, and that it must be fine-tuned to ecological conditions. We discuss the relevance of this approach to studies in animal cognition, using examples from animal foraging and decision-making, problem solving, and cognitive flexibility. Finally, we explain how even the apparent human-animal gap in sequence learning ability can be explained in terms of different fine-tunings of a similar chunking process.

**Funding:** This study was supported in part by the Israel Science Foundation grant 1126/19 to A.L., and by NSF grant FMitF-2319186, ARO grant W911NF-22-1-0061, MURI grant W911NF-19-1-0217, and a grant from the Cooperative AI Foundation to J.H.

**Acknowledgements:** We thank Yosef Prat, Ben Flickstein, two anonymous reviewers, and the guest Editor (E. Leadbeater) for useful comments that greatly improved the manuscript.

## 1. Introduction: decoding the magic

Facing the challenge of explaining how natural selection shapes diverse animal minds, it may be helpful to consider how selection shapes other diverse adaptations whose evolution is perhaps easier to understand. For example, in most textbooks of evolution, the diversification of vertebrates' limb morphology is used to demonstrate how the same set of bones has been gradually modified by natural selection to result in a variety of limbs adapted to different types of movement. This example is convincing because it is relatively easy to see how small structural modifications could gradually make some limbs better for running, jumping, swimming, flying, and so on. The physics and the mechanics involved are quite clear; there is "no magic" in how natural selection could have possibly created such remarkable adaptations.

In contrast, when it comes to cognitive adaptations, the connection between possible modifications and how they lead to improved cognitive abilities is far less clear. Despite substantial advances in the fields of animal cognition and brain sciences, recent accounts of cognitive evolution are still centered on structural adaptations, such as brain size, number of neurons, and brain connectivity (e.g., [1–3]). Indeed, there is evidence that such structural changes may be related to cognitive abilities that are considered to be advanced (e.g., [4–6]), and it is certainly reasonable to expect that advanced cognitive processes require more brain, more neurons, and better connectivity. However, it is still not clear how adding more neurons or improved connectivity generates the cognitive processes that constitute advanced cognition. In other words, if cognition is defined in terms of mechanisms that process information, construct memory representations, and make adaptive decisions based on these representations [7,8], then a theory of cognitive evolution must offer a way to understand how gradual modifications of such simple mechanisms could have resulted in diverse and advanced cognitive abilities (e.g., [9,10]; see also Leadbeater and Watrobska, this issue). To do this, and hence to allow evolutionary theory to "decode the magic" of cognitive adaptations, one must start with basic learning mechanisms whose working principles are relatively clear (e.g. simple Hebbian or associative learning), and then, try to explain everything else in brain and behavior as a consequence of their gradual modification. Given the complexity of the brain and the knowledge gaps in the field, such a theory is unlikely to be quite right [11,12]. Yet, because we don't see how a theory of cognitive evolution can bypass this challenge, we offer such a theory, which may be useful until it can be improved or replaced by better ones.

## 2. A brief sketch of the theory

We previously proposed a model of cognitive evolution that is based on the coevolution of data acquisition and learning mechanisms [13], which in turn followed some of our earlier work [14,15] and was applied to explain cognitive gene-culture coevolution [16,17], and human-like cognitive mechanisms in other animals [18]. Our model was initially just described verbally but later some of its main aspects were modeled explicitly and implemented in a set of computer simulations, demonstrating the feasibility of a gradual evolutionary change from simple associative learning, to chaining, to continuous learning where a network model of the environment is constructed [19], and finally, to a model in which chunks and hierarchies are constructed in the network [20], allowing to support features of animal creativity [21] and human language acquisition [22] (This last version successfully reproduced a range of findings in human language.) The most recent and comprehensive version of our framework is still in progress but available [23]; there we also explain how it is different from other approaches (e.g., it is not a neural-net model; see Section 8 below). Our goal here is to provide only a sketch of the theory that is needed for understanding the rest of this perspective paper. Readers seeking a brief yet explicit formulation of the core principles of the model are referred to section 3.3.3 in [23], and to the relatively simple model of the cleaner fish case in [24], which will be discussed in section 5 below.

In our framework, selection shapes cognition primarily by acting on genetic changes that affect learning and attentional (data-acquisition) mechanisms. During the animal's lifetime, these mechanisms jointly construct a network of nodes and edges (links) that represents how objects and events in the environment are associated in time and space (e.g., how they are associated with each other and with events such as finding food or encountering a predator). We therefore view the construction and the updating of the network as learning. Clearly, knowledge of certain basic associations, such as those between specific stimuli and their prescribed responses, can be innate. We view these as the initial (innate) sections of the network, which can expand significantly through learning (see Section 3.1 in [23]). The network is used for generating adaptive responses to the environment, so by enabling these responses its structure affects Darwinian fitness (see examples below as well as [19–21] for an explicit implementation in the context of animal foraging). Nodes and edges in our network are theoretical constructs; they do not correspond to neuronal cells and synapses. But very much like the concept of engrams [25], they could represent the activities or structures of communities of neurons, so their working principles should be consistent with what neuronal cells can conceivably do. The way by which external and internal signals activate the network to produce adaptive responses is discussed elsewhere [23]. For the present paper, it suffices to understand that any use of the network (regardless of how exactly it is done) depends on what information has been acquired and is thus available in the network (i.e., what has been deemed relevant by the data-acquisition mechanisms), and how it is represented by the network in terms of associations between items and their relative weight (which depends on the learning and memory parameters, as will be explained below). Clearly, in reality, the network also must be implemented by some real neuroanatomical structures and activities in the brain. However, in our framework, genetic modifications of neuroanatomical structure are favored by selection only if they release the constraints on network development, supplying the demand for storage or for better connectivity (e.g. more neurons for creating more engrams or more place cells to represent more locations). Supplying this demand may allow the network to develop, to grow, and to work in a way that produces more adaptive behaviors, but in the absence of such demand, adding neurons or increasing brain size is not expected to be adaptive (see [17] for further discussion).

While the network in our model represents the brain, there are no genes coding directly for its structure; rather there are genes controlling the basic learning and data acquisition mechanisms, thereby adjusting and fine-tuning the process of network construction. To provide a simple illustration of this process for the present paper, we start by noting that the network is built of nodes and edges that have memory weight increase and decrease parameters. These parameters, like any others in our model, may be genetically fixed, or phenotypically adjustable following some genetically based rules such as "increase weight faster under stress or upon receiving a reward" [13,23]. The evolution of these weight increase and decrease parameters, or of their phenotypically adjustable rules, is largely responsible for the evolution of cognition because these parameters determine how the network is constructed (but see explanation below regarding the complementary role of the data acquisition mechanisms). Aiming not to lose the generality of the model, we do not relate our memory weight increase and decrease parameters to specific known mechanisms of memory formation and consolidation (such as those reviewed by Leadbeater and Watrobska, this issue). However, our assumptions are in line with how such mechanisms typically behave. That is, we assume that nodes in our network can represent elements in the environment, such as *A* and *B*, if their memory weights increase whenever these elements are observed. We also assume that weights decay (decrease) unless they reach a sufficiently high level, which acts as a memory fixation threshold (in a process similar to memory consolidation or to the formation of long-term memories). Thus, elements *A* and *B* will be learned and remembered (i.e., represented in the network) only if they are encountered frequently enough to allow the nodes representing them (hereafter A and B) to increase in weight faster than they decay, until their weight crosses the fixation threshold (e.g., if the weight increases by 1 unit each time when *A* is observed and decays by 0.1 units per minute, then with a fixation threshold of 4, fixation will be reached within 30-40 minutes if *A* is observed every five minutes, but will never be reached if *A* is observed only once per hour). This learning dynamic is not only relevant to the elements, but also to their learned associations. That is, we also assume that if *B* is repeatedly experienced shortly after *A*, a directed edge leading from A to B will be created in the network, and it will have a weight that increases or decreases following the same rules. Consequently, edges in the network represent associations between elements, and their weights represent associative strength. Importantly, the weight increase and decrease parameters (and the fixation threshold) determine which elements and associations pass the test of memory decay, and are thus represented permanently in the network (as in the formation of various types of long-term memories; see Leadbeater and Watrobska, this issue), and which are eventually discarded. This test of memory decay can be viewed as a test of statistical significance that allows the animal to discard insignificant information or accidental associations, and to represent in the network only elements and associations that are likely to be real and ecologically meaningful [13,15]. It should be clear, however, that to do this successfully, the learning parameters must be tuned according to the expected distribution of elements in the environments (i.e., according to how frequently they are likely to be encountered). Otherwise, important information may decay too quickly, or background noise and misleading information may reach fixation too quickly.

To be precise, we should say that the learning parameters must be finely tuned in relation to the perceived distribution of environmental input, that is, the distribution after it passes through the animal's sensory and attentional filters (that we call collectively the *data-acquisition mechanism*). That is why we view cognitive evolution as resulting from coevolution of learning and data-acquisition mechanisms [13]. Simply put, for a stimulus to be learned it may not be sufficient that it is common in the environment (e.g., clouds in the sky), but it should also fall within the range of stimuli that the animal evolved to pay attention to (e.g., seeds-like shapes on the ground), or to assign high memory weight when highlighted by parents, social peers, or by reinforcing experiences signifying high payoffs or danger [13,15]. Note that the effect of network construction on fitness may be largely indirect. While it is easy to see how associating a particular environmental cue with food or danger is directly related to fitness, constructing a network that represents the environment is like building a cognitive map. It is difficult to predict which sections of the map may be useful in the future but constructing a good map is ultimately necessary for planning and executing adaptive courses of action [19,26].

The fine-tuning of the learning parameters becomes especially important with the transition from elemental learning (that represents elements and their associations, as described so far) to non-elemental learning. Non-elemental learning has been studied in different contexts, using different names (e.g., *configural learning* [27], *segmentation* [28], or *chunking* [29,30]). Here, we will use the term "chunking" to describe all types of non-elemental learning (as in [20,24]). The process of chunking involves the learning of combinations of elements, allowing these combinations to have a meaning that is different from that of their components (e.g., *AB* is different from *A* or *B*), or from that of similar combinations (e.g., *AAB ≠ ABB ≠ ABC ≠ AAC*), or even from that of the same combinations but in a different sequential order (e.g., *ABC ≠ ACB*; see Section 7 for an in-depth discussion of this case). To discriminate between different meanings, each chunk should be represented by a different node in the network (i.e., not merely by the nodes representing its components), so that this unique node can be associated with elements or chunks that represent its unique meaning (e.g., *AB* is associated with food while *A* and *B* are not). One way by which a chunk representing the *AB* combination can be created in our model is when a node receiving signals from the nodes representing *A* and *B* (the nodes A and B), increases in weight as a result of the simultaneous (or sequential) activation of A and B and thus becomes the AB node (see Section 7 for more details). As we will explain shortly, fine-tuning of the learning parameters becomes critical in the case of chunking because chunking involves two related computational challenges for which the solution, according to our framework, lies precisely in this fine-tuning.

The first problem is that the number of chunks that can be constructed increases exponentially with the number of learned elements, which means that, over time, a naïve approach will require increasingly (and perhaps unrealistically) large memory storage, as well as computationally challenging mechanisms for handling such enormous data sets [15,31]. The second problem (that may even kick in at a much earlier stage) is that the construction of all possible chunks from the learned elements can impair generalization and even be misleading (see Section 5 for concrete examples). Intuitively, this is like trying to learn the meaning of all possible combinations of syllables in a newly acquired language instead of trying to do so based only on the meaningful words of this language. Our model solves both problems by the dynamic process of decay and fixation. When nodes are created by random associations of elements, their weight decays quickly, but when they represent repeatedly occurring combinations, their weight increases and eventually reaches fixation (which is interestingly reminiscent of the spacing effect in memory consolidation [32]). Moreover, the fine-tuning of the memory parameters is not only critical for selecting the correct (meaningful) chunks, but by doing that, it also determines the structure of the network and the behaviors it can support. This may be easy to see when the model is applied to human language and the main chunks are words, which allows us to compose sensible and grammatically correct sentences [22]. When these parameters (and their coordination with the data-acquisition mechanism) are not well adjusted, some of the chunks can be misleading [24], or become too large and idiosyncratic, leading to a network with poor link structure that can greatly impair a wide range of cognitive performance (see [13,14,16,21] for details). In the following sections, we will show how this fine-tuning may be relevant to a range of cognitive abilities in animals, and how it may help to understand their evolution.

**3. Adaptive fine-tuning in simple forms of learning**

The idea that the fine-tuning of learning and memory parameters is consequential for learning is certainly not new. It was studied in relation to the sample size that should be used for learning [33,34], the tradeoff between exploration and exploitation [35], the length of sensitive periods in imprinting and imprinting-like processes [36,37], and it is commonly applied when optimizing learning rate parameters in reinforcement learning models (e.g., [38]). Another set of remarkable studies highlighted the tradeoffs between fast and slow learning, and short- and long-term memory under different ecological conditions [39–41]. Thus, it is already well recognized that variation in how fast and for how long animals learn and remember various associations (which in our model is determined by the weight increase and decrease parameters) can be understood in terms of adaptive tunings of the same associative learning mechanisms to different ecological conditions.

It is also well recognized that learning mechanisms may be tuned to learn certain types of information better than others, a phenomenon called *prepared learning* [42,43]. A notable example of this phenomenon is the Garcia effect, showing that rats can easily associate gastric illness with previously experienced taste but not with previously experienced light or sound [44]. The adaptive value of such tuning is that in nature, taste is far more likely to be related to the cause of the illness than light or sound. Indeed, experimental evolution in *Drosophila* has shown that manipulating the reliability of different stimulus-outcome pairs over 40 generations resulted in the evolution of greater sensitivity to the more reliable stimulus [45]. The Garcia effect also demonstrates that animals can associate events experienced hours apart rather than only seconds or minutes apart when it makes ecological sense to do that. Thus, to determine what to learn, associative learning mechanisms evolved to be tuned to certain types of sensory input as well as to some degrees of time proximities between potentially associable events. In our model, these aspects are part of the data-acquisition mechanism, which coevolves with the learning parameters, as described in Section 2 above. The examples reviewed here show that at the level of simple associative learning, it is already well established that different tunings of learning and data-acquisition mechanisms can result in diverse learning abilities and specializations.

**4. Foraging decisions: do animals need a network?**

Before delving into the complexities of chunking in networks, we should perhaps consider at which point using (or thinking in terms of) a network is necessary. Conceptually, a network model may already be useful for simple foraging decisions, implemented by two nodes representing the experience of choosing between two stimuli (e.g., opening a red or a green door), with outgoing edges leading to the experienced outcomes: finding food, or finding an empty dish (as in [19,24]). By extending such simple networks, we can easily capture the phenomenon of backward chaining, where animals associate additional stimuli with those already associated with food [46,47]. Chaining these associations itself leads to an extended network that allows the animal to navigate its way to food through a much larger cognitive space [19]. Moreover, the construction of such networks can also become intrinsically motivated, regardless of the association with the initial reward, allowing the network to grow faster and larger, which can increase foraging efficiency in structured environments [19].

But do simple foraging decisions already require a network? We believe that they do, because decisions in nature are never simple. They are usually made in some context, and animals' ability to match the context they face with a similar context experienced in the past, and predict forward, is best served by a network representation. To illustrate this point, consider a simple problem studied in house sparrows (*Passer domesticus*) that had to choose between food-related cues (sand color) when the food itself was either visible or hidden [48]. During training, sparrows in this study could learn to associate the sand color with the food setting, food amount, and handling time, all of which are informative when tested with hidden food (to predict the probability of finding food, the expected amount of food, and the cost of handling time). Indeed, when tested with hidden food, the sparrows integrated all three factors in their decisions. But when tested with visible food (seeds exposed or below a transparent cover), their choice of color was based only on how it was related to handling time, which makes sense because the probability of finding food is no longer relevant, and the amount of food is already visible. This ability to use learned associations selectively, depending on context, is not trivial, and is difficult to explain by simple conditioning. Yet, as suggested by Ben-Oren et al. [48], it can be explained if the bird matches its situation with similar situations in its memory representation, and uses this representation to predict future events. This can be done by following directed links in a network that represents past association between objects or events over time. That is, facing a feeding well covered with sand, the representation of similar situations in the past makes it possible to predict that upon searching in the sand there is some probability of finding food of a certain amount that can be eaten after some handling time. On the other hand, if the present situation is that of 6 visible seeds, the birds' starting point in the network is already at the more advanced stage from which only handling time is predicted going forward (there are no links leading from the node representing 6 seeds to "food not found" or to "2 seeds").

This network approach is also consistent with experimental evidence for context-dependent decisions in starlings that make choices based on previously experienced sequences rather than on absolute values [49], as well as with the idea that the well-known "partial-reinforcement extinction effect" [50] may be best explained by matching present situation with similar sequences experienced in the past, making it possible to predict success after failure [51] (see also [52]). Thus, animals' ability to use learned information selectively, to predict forward, and to make context-specific decisions, may already depend on their ability to construct a network. Since the content and structure of such networks are shaped by how the learning and data-acquisition parameters are finely-tuned to the environment, it is quite possible that different fine-tuning evolved in different animals and resulted in somewhat different networks, and therefore in different cognitive abilities.

**5. Chunking is easy, but the fine-tuning of chunking may not be**

The ability to create chunks is widespread among animals, as indicated by experimental evidence in mammals, birds, and insects (e.g., [53–55]). Yet, in most cases, discrimination between configurations, or between configurations and their elements, is far less accurate, and takes longer to learn, than discrimination between elements [56–58]. According to our model, it should be possible to evolve better configural discrimination by tuning the memory weight parameters so that a chunk such as AB will be created after just a few observations of the co-occurrence of A and B. However, as discussed earlier (section 2), creating chunks too fast may not be adaptive because it result in creating too many chunks that do not represent meaningful combinations. Thus, the parameters controlling chunk formation may be tuned to limit this process or to slow it down. Empirical evidence consistent with this view is, for example, that some species of *Heliconius* butterflies show better configural discrimination than other species, which may have evolved because these species have specialized to a relatively narrow range of flower types that must be distinguished from a variety of other flowers in the forest based on their unique configurations [59]. Other studies suggest that the tendency to use configural learning increases under conditions indicating its higher informative or reinforcing value, for example when a configuration such as AB is a reliable signal of receiving a reward while its isolated elements A and B signal no reward [60,61]. It has been suggested that configural discrimination (such as between rewarding AB and the non-rewarding elements A and B) may be difficult due to a tradeoff between mechanisms supporting discrimination as opposed to generalization [62,63]. We should stress, however, that this tradeoff is not only mechanistic but also functional. Even if chunking was easy and animals could have many chunks nodes in their network (in addition to nodes representing single elements), doing this would not only be costly in terms of memory storage, but would also mean that learned associations would be chunk specific. For example, an animal that already represents the chunk AB as a node in the network, and later learns that AB is associated with food (thus creating an edge leading from the node AB to a node representing food), may fail to learn that A, AC, or any other chunk that includes A is also related to food, because each of these units will be represented by a separate node that does not share the edge from AB to food. This may be maladaptive if *A*, rather than *AB*, is the signal for food, or more generally, when the element is more informative than its configural context. Indeed, theoretical analysis shows that adding chunks to a network that already represents the relationships between elements in the environment would be adaptive only when the environment is sufficiently structured, so that combinations of elements are more informative than elements alone [20]. Thus, animals' ability to succeed in non-elemental learning tasks may reflect the extent to which their tendency to chunk, or their speed of chunking, were positively or negatively selected in their natural habitat.

A particular case demonstrating why chunking must be finely-tuned in order to be adaptive is that of the cleaner fish (*Labroides dimidiatus*), which needs to learn to serve visitor clients before resident clients because visitors may leave if not served first [64,65]. This learning task, known more generally as the *ephemeral reward task*, is apparently difficult for many animals [66], and in its complex natural form cannot be solved without chunking ability [67]. That is, to learn to serve visitors (*V*) before residents (*R*), the fish must hold different representations of serving *V*, serving *R*, serving *V* then *R*, *R* then *V*, *R* then *R*, and *V* then V, so that, roughly speaking, the high reward of serving *V* first in the case of *V* then *R* (gaining two meals) will not be confused with the lower reward of doing that in the case of *V* alone, or *V* then *V* (usually ending with only one meal because the second *V* leaves). Importantly, further analysis [24] shows that having the ability to construct all these chunks is not enough because a tendency to create chunks too fast results in also creating the rarely occurring but misleading chunk RV (*R* then *V*). The creation of this chunk is possible because sometimes, especially under high client densities, after serving first a resident, the visitor who leaves is replaced by a new visitor. As a result, the cleaner may learn that choosing R first may be as adaptive as choosing V first (because it entailed the sequence R then V with a payoff of two meals). However, learning this sequence is misleading because in most cases choosing R first results in V leaving without being replaced by a new visitor. Note that learning the misleading chunk RV=2 meals is a problem only for cleaners who chunk too fast (i.e., they construct the chunk despite the rarity of the actual sequence). Thus, as a result of a tradeoff between chunking too slow and too fast, cleaner fish can solve the problem successfully only if their tendency to create chunks (determined by the chunking parameters) is finely tuned to the relative frequencies and densities of client types in their habitat [24]. This analysis may explain why only highly experienced cleaners learn to solve the problem – because chunking is selected to be relatively slow, and why cleaners' success is so variable – because both laboratory and current habitat conditions may deviate from those under which the fish evolved or developed its chunking parameters (see [24]).

**6. Generalization, problem solving, and cognitive flexibility.**

Generalization is known to be based on similarity [68]. But how is it related to the fine-tuning of chunking? One of the mechanisms that allow generalization from one item to another, even when the two are not similar (visually or acoustically), is the mechanism that makes use of their similar link structure in the network (which is how synonymous words are represented in a language model; they are associated with the same words that are likely to precede or follow them). That is, A and Z may be viewed as similar if both have edges of similar weights coming from X and Y and leading to B, C and D (Fig. 1a). This similar link structure suggests that A and Z represent things that belong to the same category (e.g., things that can be eaten, or can be climbed on, or can be passed through when navigating from X and Y to B, C, and D). They can therefore substitute for each other when planning sequences of actions based on available paths in the network leading to a goal, such as food (see Fig. 1b, and [21] for more details). This generalization process is essential for creativity and problem solving because it allows the use of previously learned associations for generating novel sequences that are not only novel, but also context-appropriate and therefore functional (as illustrated in Fig. 1c, and explicitly modeled and simulated in [21]). Intuitively, this is like taking a new combination of lines in London's underground that you have never taken before, based on familiar segments you use quite often. However, having similarity in link structure depends on how the network is structured, which in turn depends on the learning and data-acquisition parameters. If A and Z were chunked too quickly with other units in the input and are thus embedded, for example, within the chunks XA, AD, ZB, and ZD, then they do not have similar link structure and no such generalizations can be made (as illustrated in Fig. 1d, see also Fig 1 in [16] and related discussion, and explicit simulations in sections 4.6 and 4.8 in [21]). Similarly, if A and Z themselves are composed of some elements and should be chunked to be represented as such, a failure to construct these chunks can prevent the further ability to generalize across them (see Section 4.7 in [21] for an analysis of this case). Thus, by determining the structure of nodes and edges in the network, the fine-tuning of chunking can be highly consequential for animals' ability to generalize and to solve problems or innovate based on generalizations.

The same problem of network structure and generalization can greatly affect cognitive flexibility. Conventional tests of cognitive flexibility are focused on the animal's tendency to reverse preferences [69], or to inhibit a prepotent behavior in favor of another one [70], when the first behavior is no longer rewarding. Importantly, in these tests, the alternative behavior is usually available to the tested animal (e.g., reverting to the previously unrewarding key, or making a detour to reach the food). However, animals' ability to exhibit such flexibility under natural settings depends not only on their tendency to shift to a new behavior, but also on the range of alternative options offered to them (e.g., [71]). Moreover, in most cases in nature when an animal gets stuck with a behavior that is no longer successful, the alternatives are not immediately visible but must be searched for in recent or even distant memories. Even in foraging models of patch leaving it is assumed, and also demonstrated, that current diminishing reward is compared to alternative food patches in memory [72]. It should be clear therefore that animals' ability to find context-appropriate alternatives is strongly affected by how past experiences are represented in the network and can be matched with the animals' current state. For example, a representation of the sequences "A-B-C-Food" and "X-B-Z-Food" may help an animal that failed to find food after C to quickly return to B and navigate to food through Z. This action involves the flexible use of the link structure for generating the novel alternative sequence A-B-Z. On the other hand, if the sequences ABC and XBZ are represented as two distinct chunks, no such novel shortcut can be offered. As mentioned earlier, simulations of such problems demonstrate that representation that is based on large chunks promote entrenchment and rigid behavior rather than flexible behavior [21]. Moreover, it is quite possible that the tendency to shift to a new behavior (which is the flexibility measured in cognitive flexibility tests -- see above) will be shaped by experience in a way that increases this tendency in individuals whose network offered them good alternatives in the past (rewarding this tendency), and decreases it in individuals whose network could not help them to find alternatives, so they learned that the best they can do is being persistent [14].

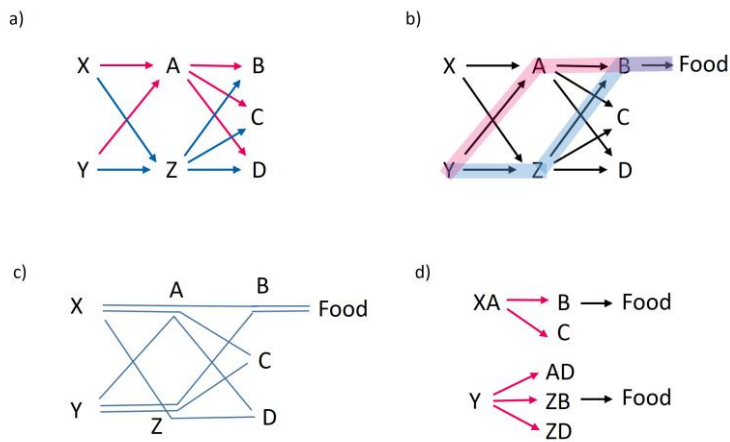

**Figure 1.** Schematic description of how similar edge structure of two nodes in the network can support generalization. **a)** The incoming and outgoing edges of nodes A (red arrows) and Z (blue arrows) have the same structure (both have incoming edges from X and Y, and outgoing edges to B, C, and D). **b)** As a result of this similar edge profile, using the network for planning a sequence of actions (by some mechanism that tracks the edges and their relative weight), makes it possible to navigate from Y to B and then to "Food", through either A (pink path) or Z (light blue path). Similarly it is also possible to navigate from X to C or to D through either A or Z (not highlighted in the figure). **c)** Although a limited set of sequences experienced by the animal (each blue line represent a particular experienced sequence) is sufficient for creating the network described in (a) and (b), the network can then allow to produce novel sequences never experienced before (and are therefore not shown in the figure but can easily visualized): Some of them can lead to "Food" and may therefore be viewed as functional (XZB-Food and YAB-Food) and some are novel but do not lead to food (XZC, XAD, YAC, and YZD). **d)** Rapid chunking may cause A and Z to become embedded within specific chunks such as XA, AD, ZB and ZD, resulting in a network that impairs generalization: the only way to reach food is by following the previously experienced sequences XAB-Food and YZB-Food, and generating any novel sequences, even non-functional ones, is impossible.

## 7. Can different tuning of chunking parameters explain the human-animal gap in sequence learning?

In an insightful review of the experimental evidence, Ghirlanda et al. [73] have pointed out that non-human animals' performance in sequence-discrimination tests is substantially inferior to that of humans. While humans may learn quickly and effectively to discriminate between stimuli ordered as AB versus BA or BAB versus BBA, non-human animals require hundreds of learning trials to start discriminating between such stimuli, and even then, their level of accuracy may not be high (e.g. 60-85% correct choices). Ghirlanda et al. suggested that humans' superior ability to learn sequential information may set us apart from other animals, and that it may be based on a different mechanism than the one used by non-human animals. They further pointed out that coding sequential information involves the problem of exponential increase in the number of possible sequences, which requires memory capacity that only the human brain may be capable of handling [74]. Therefore, according to Ghirlanda et al. [73], non-human animals do not use explicit sequential representations of stimuli as suggested above for the case of sequence-sensitive chunks (e.g., the VR chunk in the cleaner fish case). Instead, they may be using a simpler mechanism for sequence learning that is based on the different patterns of ephemeral memory traces produced by sets of stimuli that differ in their sequential order. For example, at time *t,* after hearing the sequence AB, the memory trace of B should be higher than of A because the first had more time to decay. On the other hand, when hearing the sequence BA, it should be the other way around (the B trace has more time to decay). Importantly, the trace memory model can explain why sequence discrimination in non-human animals is relatively poor, because although trace patterns of different sequences are somewhat different, they may also be quite similar (see the quantitative analysis in Ghirlanda et al. [73], where memory-trace combinations are mapped into a two-dimensional space and the distance between them determines their discriminability). Indeed, the trace-memory model predicted successfully the levels of discrimination (proportion of correct choices) of different animals tested in a wide range of sequence-discrimination tasks (see Figure 4 in [73]).

However, while the trace-memory model can explain the low proportion of correct choices in sequence-discrimination tests, it is not clear why, according to this model, it takes non-human animals hundreds of trials to learn, and how the accumulation of learning steps eventually improves performance (as illustrated by Figure 1 in [73]). According to the trace model, the discriminability of two sequences depends on how far apart the representations of their memory traces are at the time the decision is made [73], information that is already available after the first few trials. To explain why many repeated trials can be helpful, it must be assumed that memory traces are somehow coded in memory and accumulated over trials to improve discriminability. In other words, the decisions are not made directly based on short-lived memory traces, but memory traces gradually construct representations of the different sequences that will then serve the decision-making process. In our view, and as will be explained below, these sequence representations that are constructed by memory traces can be viewed as chunks, and the mechanism suggested by the memory-trace model resembles, to a large degree, the dynamics of chunk formation. Moreover, we suggest that humans' ability to learn sequential information better is not due to having a different cognitive mechanism, but can rather be captured in terms of different tunings of a similar chunking process, as we claim for many other advances in cognitive evolution.

First, it should be noted that the problem of the exponential explosion of the number of possible combinations is already relevant for any form of configural learning or chunking, even when the chunks are not sequence-sensitive. For example, an animal that encounters 20 different elements in its environment can potentially arrange them in more than a million different combinations that are not sequence-sensitive. (n different elements can be arranged in $2^n - 1 - n$ combinations [31], which is 1,048,555 when n=20). It is true that the challenge is greater for sequence-sensitive chunks, but the problem must have emerged and been solved as soon as animals used basic configural learning, which they certainly do (see Section 5). As described above, in our framework, the problem is solved by the dynamic process that keeps in memory only the chunks that pass the test of memory decay, and are therefore likely to be statistically significant and ecologically relevant [13,15,16]. That means that most potential combinations and sequences are discarded, which clearly reduces the potential memory load.

Yet how can the same chunking process result in slow and inaccurate sequence-discrimination learning in non-human animals, but still be modified to produce fast and effective sequence-discrimination learning in humans? To answer this question, we first consider the formation of chunks that are sequence insensitive, hereafter denoted {A, B} as opposed to AB and BA, where the order of the elements does matter. According to our framework (see [23] for details), the formation of the chunk {A, B} starts when the two nodes in the network representing the elements *A* and *B* (denoted here as nodes A and B respectively) are activated simultaneously (or sequentially), and there is a node n with incoming edges from both A and B (through which activation signals can be transmitted), so that the co-activation of A and B increases the memory weight of this n node (though it is still subjected to decay unless it crosses a fixation threshold – see Section 2 above). This mechanism implies that repeated co-occurrence of *A* and *B* will be represented by the weight increase, and eventually the fixation in memory, of the node n, which will then become the chunk {A, B}. The activation signals sent from A and B to the node n may resemble the memory traces of *A* and *B* in the memory-trace model [73]. However, in our framework they have the additional function of turning into lasting memories by gradually increasing the memory weight of the chunk {A, B}.

As described so far, the chunk {A, B} represents the statistically significant co-occurrence of *A* and *B*, regardless of their sequential order of appearance. It may represent simultaneous (or nearly simultaneous) co-occurrence if the levels of activation decay sharply, or a broader range of time proximity if these levels decay slowly (see Fig. 2a). But it cannot represent sequential order, because changing the order would not change the outcome (see Fig. 2b). To discriminate between *AB* and *BA*, two different nodes must be created, one for each sequence. Creating a node that represents *AB* but not *BA* is possible when there is a node n' in the network that is positioned in a way that causes the signal from the node A to take a bit longer to arrive than the signal from node B. We can think of it as if the signal coming from A travels a larger distance before reaching n' (see Fig. 2c), though the actual mechanism may be more complicated (e.g., it may go through an intermediate node). Regardless, with this architecture, the level of activation of n' will increase most if *A* is observed somewhat before *B*, because then the two signals will arrive together. Similarly, there might be a node n'' for which the signal from B takes a bit longer to arrive, causing its activation level to increase most in response to the input *BA* (see Fig. 2c). Since an increase in activation level leads to an increase in weight, nodes n' and n'' will represent the different occurrence of *AB* and *BA*, respectively.

It is important to note that accurate discrimination between *AB* and *BA* can be achieved only when the structure of the network and the activation patterns are adjusted to result in sequence-specific activation, and consequently in sequence-specific chunks, as illustrated in Fig. 2c. This is unlikely to happen by chance, and may become possible only after these parameters are finely tuned to that aim by natural selection. This specialization is probably what we see in humans, for whom language and other cultural innovations made sequential learning critically important [16,74]. On the other hand, this may not be the situation in most other animals, where sequence sensitivity is less important (at least in the contexts reviewed by Ghirlanda et al. [73]), and has the down side of impeding generalization (see Section 5). Thus, for most animals, the network structure and the activation patterns may look as described in Fig. 2d, and result in little sequence sensitivity (at least initially). Little sequence sensitivity means that both *AB* and *BA* cause peak activations in all the three nodes (n, n' and n''). However, as we explain below, small differences in these activation levels can still make it possible to gradually improve sequence sensitivity after many learning trials, as observed in animal experiments [73].

We demonstrate this process in Fig. 2d. Under the conditions described in Fig. 2d, activation levels do not decay very fast, and the delay in the arrival of the signal of the second stimuli to n' or to n'' is relatively short. Consequently, the activation of A and B by the sequence *AB* results in activation peaks in all three nodes, with small differences between them (activation peak of n'>n>n''; see Fig. 2d). The same happens with the sequence *BA*, but in the opposite direction (with activation peak of n''>n>n', not shown in Figure 2d because it is simply the mirror image). The fact that in response to both sequences (*AB* and *BA*) all three nodes are activated to some degree, implies that any discrimination mechanisms that make use of these activation levels is likely to make some errors (as also expected by the memory-trace model). However, in the chunking model, repeated encounters increase the weight of the nodes, and decisions are made based on these relative weights. Importantly, repeated encounters with the sequence *AB* increases the weight of node n' faster than that of the other nodes because in each encounter, the node n' is activated slightly more than the others (see Fig. 2d). The same will happen with the sequence *BA* and the node n''. Thus, after many learning trials, the small differences in activation can gradually build up larger differences in weight that can facilitate better discrimination (because higher weight increases the associative strength with reward, so that if *AB* is rewarded, the node n' that represents *AB* will be associated more strongly with a reward, and therefore preferred). This process may initially be slow because weights also decay after each observation until they cross a fixation threshold (see above), but this slow rate of sequence learning is precisely what we see in animal experiments (Fig. 1 in [73]), as well as in the cleaner fish case discussed in Section 5 above.

In humans, on the other hand, the specialized AB and BA chunks can be created much faster. First, because the activation peaks of n' and n'' are initially stronger due to the adjusted synchrony in time of arrival (see Fig. 2c). Second, it is also possible that in humans, the weight-increase parameters have evolved to increase the weight faster under contexts indicating the need for sequential learning. Expressing this high-level model at the neuronal level is still premature, but recent evidence for cortical neuronal cells in humans and macaques representing different solutions for efficiency-robustness trade-offs [75] may be worth exploring further in this context.

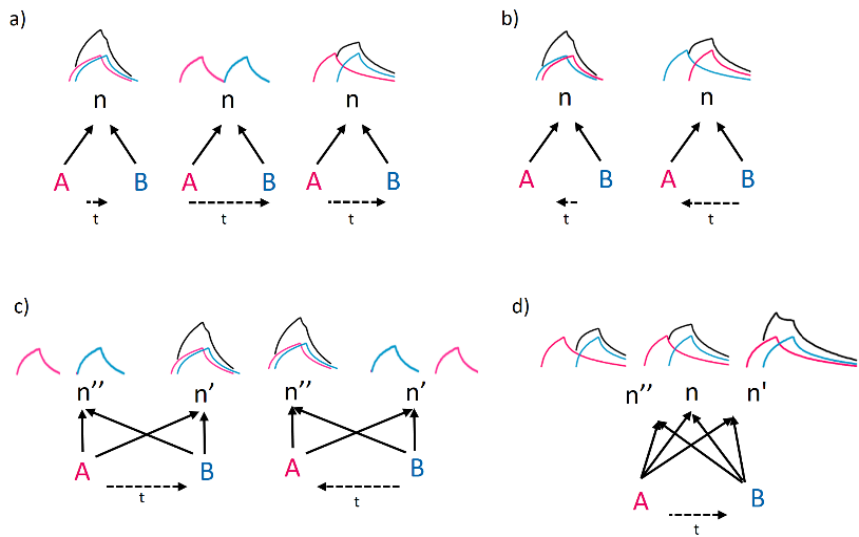

**Figure 2.** Schematic description of the process that creates chunks that are either sequence insensitive (cannot represent the sequential order of *A* and *B*) or sequence sensitive (that represent this sequential order). A and B are nodes representing the elements *A* and *B*. Solid arrows are edges leading from A and B to n, n' or n'', which are the nodes that will form the chunks {A, B}, AB, and BA, respectively (see text for full description). Horizontal dashed arrows indicate whether the sequence is from *A* to *B* or from *B* to *A*, and the length of the dashed arrow indicates the time difference between the initiations of each event. The red and blue curves represent the activation profiles of the nodes A and B, respectively, while the black curves represent the sum of their activation when these activations overlap in time, and may thus create an activation peak. The activation profile of each element is assumed to increase during its occurrence and decrease (decay) as soon as it ends. **a)** The formation of a sequence-insensitive chunk {A, B} is expected when the edges leading from A and B are symmetric, so the time needed for the activation signals to reach n is the same. Repeated peak activations will then create the chunk {A, B}, either when *A* and *B* occur simultaneously or nearly simultaneously (left) or one after the other if the decay is slow (right), but not when the activation of the first decays before the activation of the other (centre; no overlapping activation). **b)** Reversing the sequential order from *AB* to *BA* does not change the peak activation curves (compare with (a)), meaning that the chunk n is sequence insensitive. **c)** When the edges leading from A and B are asymmetric, causing the signal from A to arrive later to n' and the signal from B to arrive later to n'', sequential activations of A then B versus B then A can result in different peak activations of n' and n'', respectively, which after repeated occurrences create the sequence-sensitive chunks AB (formerly n') and BA (formerly n''). **d)** When the asymmetry in signal-arrival time is less pronounced than in (c), and the decay is slower, the sequence *AB* results in peak activations in all three nodes, though this peak is higher and broader in n' (see text for further explanations and implications).

## 8. Conclusions and outlook

Using a theoretical framework developed with our colleagues over the past fifteen years, this perspective paper aims to highlight the critical role of fine-tuning of learning and chunking mechanisms in the process of cognitive evolution. Although cognition is increasingly recognized as reliant on memory representations and cognitive maps organized in complex networks (e.g., [25,26]), the evolution of cognition has predominately been examined in terms of traits, abilities, or brain morphology, rather than through the developmental processes that construct these networks via learning. Our approach, on the other hand, focuses specifically on the evolution of these developmental processes, addressing some of the major challenges in constructing a network capable of representing the environment reliably and effectively, so it can support the production of adaptive behaviors.

Examining several lines of evidence, we argue that the main challenge in network construction is not merely to evolve the "ability" to create associations and chunks in the network, but rather to finely-tune these abilities to ecological conditions, so that only associations and chunks that are likely to be useful for producing adaptive behaviors will be learned and remembered. This fragile dynamic of fine-tuning may explain why natural selection acts to limit learning speed and memory capacities, and why it may do so to different degrees in different species. It may also explain, as discussed above, how gradual modifications of chunking parameters can lead to seemingly major evolutionary innovations, such as humans' exceptional sequence-learning ability (see [16,17] for how these learning dynamics may influence cognitive evolution in response to cultural innovations).

Finally, as we acknowledged earlier, our modeling approach may not be entirely accurate. However, we believe that the core concept of fine-tuning must be inherent to cognitive evolution. One major goal for future work will be to refine and test our model in light of emerging neurobiological research, which may become feasible in the years ahead. Another goal is to clarify the relationship between our model and deep neural-net models, which enable Large Language Models (LLMs) to exhibit human-like cognitive abilities. Comparing these two approaches is not as straightforward as it might seem. Despite the remarkable success of LLMs, their internal processes remain partially opaque, and their resemblance to biological brains is still debated [76-78]. Nevertheless, the two approaches differ fundamentally in at least two major ways: First, our model applied memory limitations and slow learning, which characterize biological brains, to fine-tune network construction and manage computational challenges. In contrast, LLMs impose no such memory constraints, relying on different mechanisms for network construction and fine-tuning. Second, in our model, the process of learning and chunking is aimed at creating nodes in the network that explicitly represent elements or combinations of elements in the real world, such as objects or words. LLMs, on the other hand, lack an explicit process for forming symbolic representations of objects or words, and whether such representations emerge within their networks has only recently been explored [79, 80]. It is possible that at some level, similar processes or working principles operate in both models, but further research is needed to determine the extent of this similarity.